# Stability of spherical stellar systems II : Numerical results

Jérôme Perez [1,2], Jean-Michel Alimi[3], Jean-Jacques Aly [1], Hans Scholl[4]

[1] *CEA, DSM/DAPNIA, Service d'Astrophysique (CNRS URA 2052) - CE Saclay - 91191 Gif sur Yvette Cedex - France*
[2] *ETCA/CREA - 16 bis av. Prieur de la côte d'or - 94114 Arcueil Cedex - France*
[3] *Laboratoire d'Astronomie Extragalactique et de Cosmologie - Observatoire de Meudon - 5, Place Jules Jansen - 92 195 Meudon - France*
[4] *Département Cassini - Observatoire de la Côte d'azur - BP 229 - 06 304 Nice Cedex 4 - France*

**ABSTRACT**
We have performed a series of high resolution N-body experiments on a Connection Machine CM-5 in order to study the stability of collisionless self-gravitating spherical systems. We interpret our results in the framework of symplectic mechanics, which provides the definition of a new class of particular perturbations: The preserving perturbations, which are a generalization of the radial ones. Using models defined by the Ossipkov-Merritt algorithm, we show that the stability of a spherical anisotropic system is directly related to the preserving or non-preserving nature of the perturbations acting on the system. We then generalize our results to all spherical systems.

Since the "isotropic component" of the linear variation of the distribution function cannot be used to predict the stability or instability of a spherical system, we propose a more useful stability parameter which is derived from the "anisotropic" component of the linear variation.

**Key words:** Stellar dynamics – celestial mechanics – instabilities

## 1 INTRODUCTION

A steady state of a collisionless self-gravitating system – in which the distribution function (DF hereafter) is determined only by the isolating integrals of motion (Jeans' theorem) – may be subject to some form of dynamical instability. After being slightly perturbed, it thus suffers a rapid evolution, which may lead to significant changes after a few crossing times. It is well-known, for instance, that spherical anisotropic systems with components moving mainly on radial orbits, are unstable. The physical origin of this instability has been extensively investigated analytically ((Antonov 1973), (Palmer & Papaloizou 1986)) and numerically ((Henon 1973), (Barnes et al. 1986), (Merritt & Aguilar 1985), (Dejonghe & Merritt 1993), (de Zeeuw & Franx 1991)). In particular, Fridman and Polyachenko (Fridman & Polyachenko 1984) reached the conclusion that a spherical system is unstable once a parameter $\xi$ – actually equal to the ratio between the radial and transversal kinetic energies of the system – exceeds some critical value ($\xi > 1.7 \pm 0.2$). However, this criterion is only an empirical suggestion, and unfortunately, subsequent simulations ((Palmer 1994a), (Palmer 1994b)) have shown its invalidity in many cases. It thus appears at present that there are no simple and general stability criteria in which intervenes a single parameter depending on the state of the gravitational system.

In this paper, we revisit this problem numerically, and propose a stability parameter for all spherical anisotropic systems. Our work rests on some analytical results ((Perez 1995), (Perez & Aly 1995a), Paper I hereafter) which have been recently obtained by working in the framework of the symplectic formulation of Vlasov-Poisson equations (for an introduction to the symplectic formalism, see, e.g., (Arnold 1978)). In this approach – first introduced in the gravitational context by Bartholomew and Kandrup ((Bartholomew 1971), (Kandrup 1990), (Kandrup 1991a)) and thus developed by Perez and Aly ((Perez 1995), (Perez & Aly 1995b)) – the Vlasov-Poisson system is rewritten in algebraic form, from which one can get a formal expression giving the value at any time of an arbitrary functional of the DF (whose initial value is assumed to be given). For a steady state system submitted to a "symplectic" perturbation induced by a "generator" $g$, it is then possible in particular to write a Taylor expansion of its energy in terms of $g$, and to deduce linear stability criteria based on the consideration of the first non-vanishing term in that expansion. Using this general result, Perez and Aly (Paper I) demonstrated some form of stability of the anisotropic spheres against the so-called "preserving perturbations" (whose class contains as a small subclass all the radial perturbations), thus providing a generalization of an earlier result (see below). Clearly, this is not sufficient to give a complete description of the stability properties of a spherical system (except in the two limiting cases of a stable isotropic system and of an unstable one with purely radial orbits), and it was just conjectured in Paper I that the stability in the intermediate cases is directly related to the preserving or non-preserving nature of the perturbations. The numerical simulations presented here seem to confirm this guess.

It may be worth recalling that two other analytical approaches have been developed in the last thirty years for studying the stability of collisionless self-gravitating systems. The first one rests on an angle-action description and on a decomposition of the gravitational potential into normal modes, and it applies to a large class of systems ((Fridman & Polyachenko 1984), (Goodman 1988), (Palmer & Papaloizou 1986), (Polyachenko 1988)). However, the calculations and the derived stability criteria are often tedious, and difficult to apply. The second method, inspired from plasma physics (Laval et al. 1965), is related to the variational energy principle, and it has been adapted by Kulsrud and Mark (Kulsrud & Mark 1970) to gravitational systems. This method has yielded interesting criteria in two situations. On the one hand, the stability of radially perturbed isotropic or anisotropic spheres has been obtained – the most general result here being due to Doremus and his collaborators (Doremus et al. 1971), whose analytical arguments were thus considerably simplified by (Sygnet et al. 1984) and (Kandrup & Sygnet 1985). On the other hand, the stability of isotropic spheres against non-radial perturbations has also been established. The proof here appeals to the well-known Antonov-Lebovitz theorem (Binney & Tremaine 1987) on the stability of gaseous spheres ((Antonov 1973), (Lynden-Bell & Sannit 1969)), a simplified derivation of which has been reported most recently by Aly and Perez (Aly & Perez 1992). A strong limitation of the energy principle arises from the necessity to have the "operator of the small motions" being Hermitian – which is not the case, e.g., for a non-radially perturbed anisotropic sphere.

The paper is arranged as follows. In Section 2, we outline the main analytical results obtained in the framework of symplectic mechanics concerning the stability of collisionless self-gravitating spherical systems. We present in Section 3 the numerical methods for simulating the initial conditions and the dynamical evolution of such systems. In Section 4, we give a physical interpretation of their (in)stability in terms of the "symplectic quantities" introduced in Section 2. Finally (Section 5), we deduce from the numerical results two laws concerning the stability of any anisotropic collisionless self-gravitating spherical systems.

## 2   THEORETICAL CONTEXT AND ANALYTICAL RESULTS

If $f$ is the DF of the system in the $\{\xi := (\mathbf{x}, \mathbf{v})\}$ phase-space, $\psi(\mathbf{x})$ the gravitational potential and $G$ the gravitational constant, the evolution of the system (during a period compatible with the collisionless assumption) is given by the Vlasov-Poisson system

$$\begin{cases} \dfrac{\partial f}{\partial t} + \mathbf{v}.\nabla_x f - \nabla_x \psi . \nabla_v f = 0, \\ \psi(\mathbf{x},t) = -4\pi G \displaystyle\int \dfrac{f(\mathbf{x}',\mathbf{v}',t)}{\mid \mathbf{x}-\mathbf{x}' \mid} d\xi', \end{cases} \qquad (1)$$

where $d\xi := d\mathbf{x}d\mathbf{v}$ denotes an infinitesimal volume of phase space.

It is well known that Vlasov's equation (1) can be written in the standard Poisson Bracket form

$$\frac{\partial f}{\partial t} = -\nabla_v E . \nabla_x f + \nabla_x E . \nabla_v f =: \{E, f\}, \qquad (2)$$

where $E = \mathbf{v}^2/2 + \psi$ is the one particle energy of the system. $E$ can be viewed as the generator of the canonical transformation describing the motion and its conjugated quantity is the time $t$. More generally, the evolution equation of $f$ in any transformation defined by a generator $g_1$ and its conjugated quantity $\lambda$ can be written

$$\frac{\partial f}{\partial \lambda} = \{g_1, f\}. \qquad (3)$$

The previous equation can also be generalized to yield an evolution equation for any functional of $f$. The time-evolution is then generated by the total physical energy $H$ associated with $f$

$$H[f] = \int \frac{|\mathbf{v}|^2}{2} f(\mathbf{x},\mathbf{v},t) d\xi - \frac{G}{2} \int d\xi \int d\xi' \frac{f(\mathbf{x},\mathbf{v},t)f(\mathbf{x}',\mathbf{v}',t)}{\mid \mathbf{x}-\mathbf{x}' \mid}. \qquad (4)$$

$H$ is a functional of $f$ and its functional derivative is

$$\frac{\delta H[f]}{\delta f} = E. \qquad (5)$$

For any $F[f]$ we have

$$\dot{F}[f] := \frac{\partial}{\partial t} F[f] = \int \frac{\delta F}{\delta f} \dot{f} d\xi. \qquad (6)$$

Inserting the bracket form of Vlasov equation (2) and equation (5) into equation (6), we get after an integration by part the functional form of Vlasov's equation describing the time-evolution of $F[f]$ (Morrisson 1980)

$$\dot{F}[f] = \int f \left\{ \frac{\delta F}{\delta f}, \frac{\delta H}{\delta f} \right\} d\xi := [F, H][f]. \qquad (7)$$

$[,]$ is a Lie bracket.

More generally, the functional evolution equation of $F[f]$ for any transformation defined by a functional generator $G_1(f) := \int g_1 f \, d\xi$ can be written

$$\forall F \quad \frac{\partial}{\partial \lambda} F[f] = \int f \left\{ \frac{\delta F}{\delta f}, \frac{\delta G_1}{\delta f} \right\} d\xi = [F, G_1][f]. \tag{8}$$

The solution to (8) can be written in exponential form as

$$\forall F \quad F[f] = (e^{[\cdot \, , \, G_1] \lambda} F)[f_{\lambda_o}], \tag{9}$$

or more explicitly as a Taylor expansion,

$$\forall F \quad F[f] = F[f_{\lambda_o}] - [G_1, F][f_{\lambda_o}](\lambda - \lambda_o) + \frac{[G_1, [G_1, F]][f_{\lambda_o}]}{2!}(\lambda - \lambda_o)^2 - \frac{[G_1, [G_1, [G_1, F]]][f_{\lambda_o}]}{3!}(\lambda - \lambda_o)^3 + \cdots, \tag{10}$$

where $f_{\lambda_o}$ denotes the value of the DF at the point $\lambda = \lambda_o$.

In the particular case $F = H$, equation (10) provides a development of the total energy of the system. When $f_{\lambda_o}$ is an equilibrium DF $f_o$, the first non vanishing term of this development gives a linear stability criterion for the system, independently of its geometry.

$$H^{(1)}[f_o] = -[G_1, H][f_o] = -\int f_o \left\{ \frac{\delta G_1}{\delta f}, \frac{\delta H}{\delta f} \right\} d\xi = \int g_1 \{E, f_o\} d\xi \tag{11}$$

is clearly vanishing ($\{E, f_o\} = \dot{f}_o = 0$), while

$$H^{(2)}[f_o] = \frac{[G_1, [G_1, F]][f_o]}{2!} \tag{12}$$

can be written after some straightforward algebra ((Perez 1995), Paper I)

$$H^{(2)}[f_o] = -\frac{1}{2} \int \{g_1, E\}\{g_1, f_o\} d\xi - \frac{1}{2} \int \frac{\{g_1, f_o\}.\{g'_1, f'_o\}}{|\mathbf{x} - \mathbf{x}'|} d\xi d\xi'. \tag{13}$$

The positiveness of $H^{(2)}[f_o]$ determines the *stability* of the system against perturbations generated by some $g_1$. Using a less general method, this result had been previously obtained by Bartholomew (Bartholomew 1971) and Kandrup ((Kandrup 1990),(Kandrup 1991a)). It is important to notice that this criterion on the positiveness of $H^{(2)}[f_o]$ is only a stability criterion, it tells us nothing about the linear *instability* of the system. Kandrup (Kandrup 1991b) showed that, if $H^{(2)}[f_o]$ is negative, the system develops a secular instability in the presence of dissipation.

In the particular case of spherical self-gravitating systems, where the equilibrium DF depends only on the one-particle energy $E$ and squared total angular momentum $L^2$, the linear variation of the DF, $\delta f := \{g_1, f_o\}$, which appears in the second order energy variation (equation (13)), can be written

$$\delta f = \frac{\partial f_o}{\partial E} \{g_1, E\} + \frac{\partial f_o}{\partial L^2} \{g_1, L^2\}. \tag{14}$$

When one considers a preserving perturbation defined by $\{g_1, L^2\} = 0$, an anisotropic spherical system behaves as a stable isotropic spherical system [*], with $H^{(2)}[f_o]$ positive ((Perez 1995), Paper I). On the contrary, systems with components evolving only on radial orbits are unstable. Such systems can be described by an equilibrium distribution function $f_o(E, L^2) = \nu(E)\delta(L^2)$ where $\nu(E)$ is an arbitrary monotonic function of $E$, and $\delta(L^2)$ denotes a Dirac distribution for $L^2$. In this case, it is then easy to show that all preserving physical perturbations are vanishing. The two previous opposite cases (stable isotropic spherical systems and unstable radial-orbit anisotropic spherical systems) suggest that the stability of a spherical self-gravitating system is directly related to the preserving or non-preserving nature of the perturbations acting on the system. In order to confirm this conjecture in all the intermediate cases, which cannot be studied analytically, we have performed numerical simulations. Moreover, under the light of previous analytical work, we will propose also a stability parameter.

### 3 NUMERICAL METHODS

#### 3.1 Initial conditions

We use the Ossipkov-Merritt algorithm ((Ossipkov 1979), (Merritt 1985a), (Merritt 1985b), (Binney & Tremaine 1987)) which is well adapted for generating anisotropic self-gravitating spherical systems with various physical properties (see Appendix).

This algorithm consists to deform the isotropic density $\rho_{iso}(r)$ deduced from a given isotropic gravitational potential $\psi_{iso}(r)$ in the following way ($r = |\mathbf{x}|$):

$$\rho_{ani}(r) := \left(1 + \frac{r^2}{r_a^2}\right) \rho_{iso}(r). \tag{15}$$

---

[*] We consider all along this paper only systems with a DF which admits a monotonous decreasing dependence with respect to all the isolating integrals of motion ($\frac{\partial f_o}{\partial E} < 0$, and $\frac{\partial f_o}{\partial L^2} < 0$).

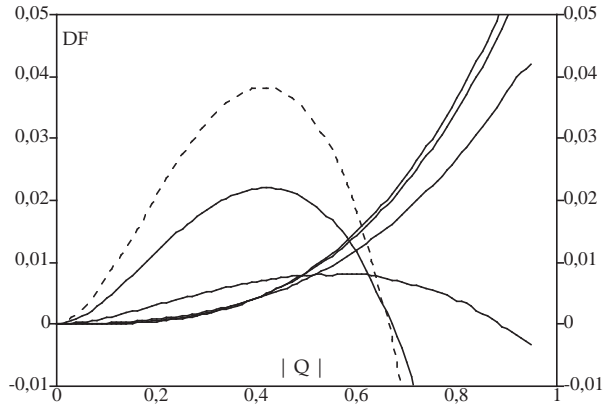

**Figure 1.** Phase space DF for $n = 4.5$ with $r_a = 0.75(\text{dot}), 1, 2, 5, 10, 100$, respectively. As $r_a$ decreases, $f$ becomes progressively smaller at low energy ($Q \longrightarrow 1$).

The anisotropic radius $r_a$ controls the deformation. Using the Abel inversion technique, this procedure allows us to generate an anisotropic DF depending both on $E$ and $L^2$ through the variable $Q := E + L^2/2r_a^2$:

$$f_o(Q) = \frac{\sqrt{2}}{4\pi^2} \frac{d}{dQ} \int_Q^0 \frac{d\psi_{iso}}{\sqrt{\psi_{iso} - Q}} \frac{d\rho_{ani}}{d\psi_{iso}}. \tag{16}$$

The velocity anisotropy at any radius $r$ is given by

$$\frac{\sigma_r^2}{\sigma_t^2} := \frac{<v_r^2>}{\frac{1}{2}<v_t^2>} = 1 + \frac{r^2}{r_a^2}. \tag{17}$$

The model is isotropic near the center and becomes more and more anisotropic outwards.

For the isotropic potential, we have chosen here the polytropic model. Thus $\psi_{iso}$ is a solution to the Lame-Emden differential equation

$$\frac{1}{r^2}\frac{d}{dr}\left(r^2\frac{d\psi}{dr}\right) = (-1)^n 4\pi G c_n \psi^n, \quad c_n = \frac{(2\pi)^{\frac{3}{2}}\Gamma(n-\frac{1}{2})}{\Gamma(n+1)}, \tag{18}$$

whith $\psi(0)$ being taken as a free parameter. We have set $\psi(0)$ equal to $-1$. Because of the spherical symmetry, we have $d\psi/dr = 0$ at $r = 0$.

The system admits a finite density and a finite mass provided the polytropic index $1/2 < n \le 5$ (Chandrasekhar 1957). The total mass of the system is fixed equal to 1 in our numerical simulations. Tuning $n$ and $r_a$, we can then modify the only free physical parameters of the system: The size and the dynamical time. The size increases as $n$ for a given $r_a$, while the dynamical time increases as the anisotropy of the system for a given $n$. All the models are virial-relaxed ($\eta = 2E_{kinetic}/E_{potential} = -1$). One family of DF's are plotted on Fig. 1. The DF generated from such a procedure do not present any fixed-point as in the Fig. 2a of Merritt (Merritt 1985a). As a matter of fact, since we have not constrained the size of the system, the gravitational potential depends in our case on $r_a$ for a given $n$ (Fig. 2). Consequently, there is no more fixed-point.

In order to set up the initial conditions of our N-body numerical simulations, we now randomly choose the positions and the velocities for $N$ particles from the DF. We plot in Fig. 3 the radial density profile and in Fig. 4 the velocity anisotropy deduced from one simulation of the system with $n = 4$ and $r_a = 2$. The agreement between our initial numerical conditions and the theoretical models are fully satisfactory.

The Ossipkov-Merrit models admit a fundamental limitation. As a matter of fact, for a given polytropic index $n$, there does exist a critical value of $r_a$ for which the DF becomes negative and unphysical in some region of the phase space (Fig. 1). Merrit (Merritt 1985a) interprets this limitation as a simple illustration of the well-known fact that an arbitrary spherical mass distribution cannot always be reproduced by radial orbits. We have nevertheless considered models with $r_a = 1$ or $r_a = 0.75$ for example, which admit a negative DF in a region of phase space (figure 1). In these cases, however, we have arbitrarily set the DF equal to zero in this region, the positions and the velocities of the particles being thus choosen only after this truncation. As we shall see hereafter, our conclusions concerning the stability or the instability of such systems with a strong anisotropy have been obtained in the same way as for the realistic Ossipkov-Merrit systems with $r_a \ge 2$.

Finally, since each particle is initialized independently, the equilibrium DF $f_o(E, L^2)$ of the system is in fact slightly perturbed. The perturbation is due to local Poissonian fluctuations of the density. The dynamical evolution of the system represents then the response of an anisotropic self-gravitating spherical equilibrium system submitted to such a perturbation.

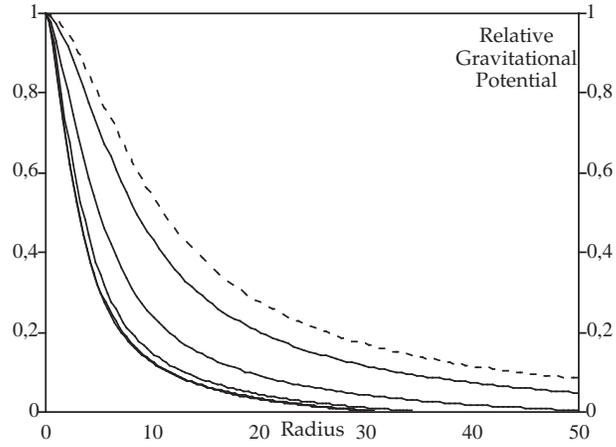

**Figure 2.** Relative Gravitational Potential for $n = 4.5$ with $r_a = 0.75 (dot), 1, 2, 5, 10, 100$, respectively. The radius $R$ at which $\psi$ vanishes, increases with the anisotropy, $(r_a \longrightarrow 0)$.

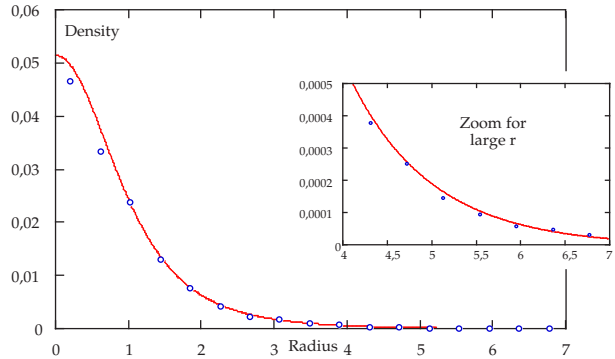

**Figure 3.** Radial density profile for a system with $n = 4$ and $r_a = 2$. The solid line is derived from equations (16) and (18), the symbols correspond to the radial density profil deduced from particles.

### 3.2 Numerical integrations

Particles of mass $m_i$ and $m_j$ interact through the softened potential

$$\phi_{ij} = Gm_i m_j/(r_{ij}^2 + \epsilon^2)^{1/2}, \tag{19}$$

where $r_{ij}$ is their separation. The softening parameter $\epsilon$ is essentially a particle radius. In our simulations, $\epsilon = 0.05$. The most obvious algorithm to integrate equations (19) is direct summation ((Aarseth 1972), (Aarseth 1985)). Unfortunately, the number of operations in this algorithm grows as $N^2$. On classical sequential computers, it is at present not possible to consider a large number of particles ($N \leq 10000$) and to perform many numerical simulations. On the other hand, the direct summation algorithm is particularly adapted to massively parallel computers like a Connection Machine. In this paper, numerical simulations have been performed on a Connection Machine CM-5 by using a "Digital Orrery" algorithm, first developed on a Connection Machine CM-2 ((Hillis & Barnes 1987), (Alimi & Scholl 1993), (Serna et al. 1994)). In this case, one physical or virtual processor is assigned to each particle. One can then imagine two rings of processors containing both the same set of N-body coordinates. One ring starts to turn around stepwise by the use of CM instructions. At each step, forces between all adjacent couples of bodies placed on the two different rings are calculated in parallel. A complete force calculation between all particles requires $N - 1$ such steps. To illustrate the computational requirements, for a simulation with 8192 particles, the main part, computing all interacting forces between all particles, took less than 10 seconds per time-step on a 32-nodes CM-5. The accuracy of the integration can be illustrated by the variation of the total energy throughout a complete calculation, typically 30 dynamical times resolved by 10000 time-steps. This variation was smaller than 0.01 per cent. Hereafter, the set of numerical simulations performed have been made with 8192 particles. In order to get statistics to put error bars on our results, each model (a couple $n$, $r_a$) has been done 6 independent times. Some experiences with more particles have been performed (65536) without significant changes in comparison with the works presented here.

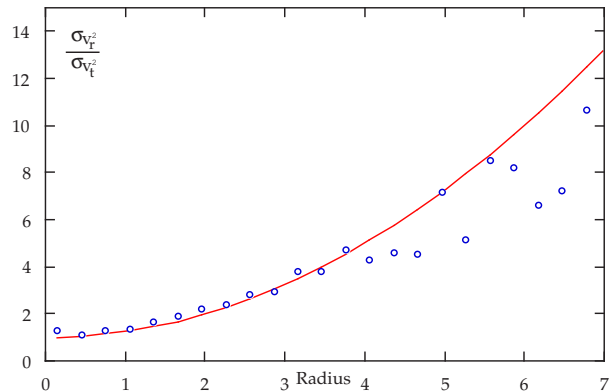

**Figure 4.** The radius dependence of the ratio between the radial velocity dispersion and tangential velocity dispersion for $n = 4$ and $r_a = 2$. The plain line corresponds to equation (18), and the symbols correspond to the same quantities deduced from the particles.

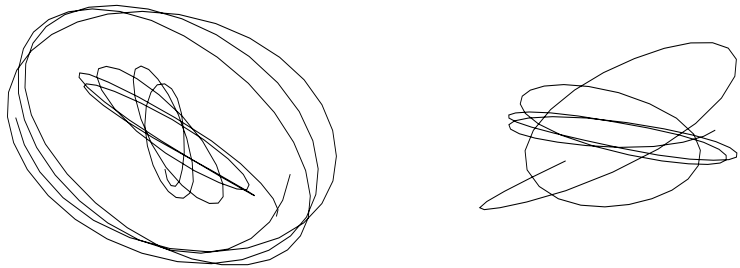

**Figure 5.** Left diagram : The orbits of two particles in a stable case $n = 4$, $r_a = 100$ we can see the regular precession due to the angular momentum and the sphericity of the potential. Right diagram : By opposition, in an unstable $n = 4$, $r_a = 0.75$ case the orbit of a particle with low angular momentum can be trapped.

## 4 NUMERICAL RESULTS

### 4.1 Morphological (in)stability

The physical mechanism of radial-orbit instability for collisionless self-gravitating systems is well-known. It is described in detail by several authors ((Antonov 1962), (Palmer 1994b)). The morphological deformation resulting from such an instability is mainly due to the trapping of particles with a low angular momentum in a bounded area of space. We can see such trapping in our two opposite stable ($n = 4$, $r_a = 100$) and unstable ($n = 4$, $r_a = 0.75$) cases on figure 5 To evaluate this deformation of an initial spherical system into an ellipsoid, it is convenient to use the axial ratio defined from the moment of inertia tensor $I$ (Allen et al. 1990). From the three real eigenvalues of $I$, $\lambda_1 \leq \lambda_2 \leq \lambda_3$, we compute the axial ratios $a_1 = \lambda_1/\lambda_2$ and $a_2 = \lambda_3/\lambda_2$. These two quantities, which can always be defined because an eigenvalue does never vanish, satisfy $a_1 \leq 1 \leq a_2$.

In Figs 6, 7 and 8, we plot the evolution of both the axial ratios and the virial ratio $\eta$ for three classes of systems defined by 3 different polytropic indexes $n$. In each class, we have considered 6 different models which range from a purely radial model with $r_a = 0.75$ to a quasi isotropic one with $r_a = 100$. A large range of systems with various physical properties (size, dynamical time....) (see Appendix), are thus taken into account in the set of these simulations.

All systems are initially spherical and virial-relaxed. $a_1$ and $a_2$ are equal to 1, and $\eta$ is equal to $-1$ (see section 3.1). The temporal average value for $\eta$ along the evolution is also equal to $-1$. However, during the first steps, at the same time when the systems possibly deforms, a significant fluctuation of $\eta$ appears clearly. It disappears later, $\eta$ staying equal to $-1$. Independently of the value of the polytropic index $n$, weakly anisotropic systems ($r_a > 2$) keep their initial spherical geometry during the evolution. Their axial ratios does never differ significantly from unity ($a_1 \simeq a_2 \simeq 1$). They are morphologically stable. On the contrary – but still independently of $n$ – systems with a low initial anisotropic radius $r_a \leq 2$. deform inevitably, reach a new non-spherical configuration, and stabilize in a new ellisoidal configuration ($a_1$ or $a_2$ are significantly different from unity). This is exactly the effect of the radial orbit instability, which flattens the system (see ((de Zeeuw & Franx 1991) for a review). The collisionless hypothesis is fundamental for interpreting our results. Consequently, we do not have continued our numerical simulations beyond a few hundred dynamical times in order to avoid a later evolution where two-body relaxation arises. However, the totality of our models reach a new relaxed state before 30 $T_d$. We thus present our results for this interval. Moreover, it is important to notice on figures 6, 7 and 8 the evolution of the virial ratio. This parameter, which stabilize also at the $-1$ value before 30 $T_d$, is a good indicator of the dynamical state of the system, which presents all the garanties of no further dynamical evolution. Previous studies, which make no use of this parameter, must illustrate their results on much larger time scales. In the next section, we interpret all these numerical results with the help of the symplectic formalism described in Section 2. We also propose a stability parameter.

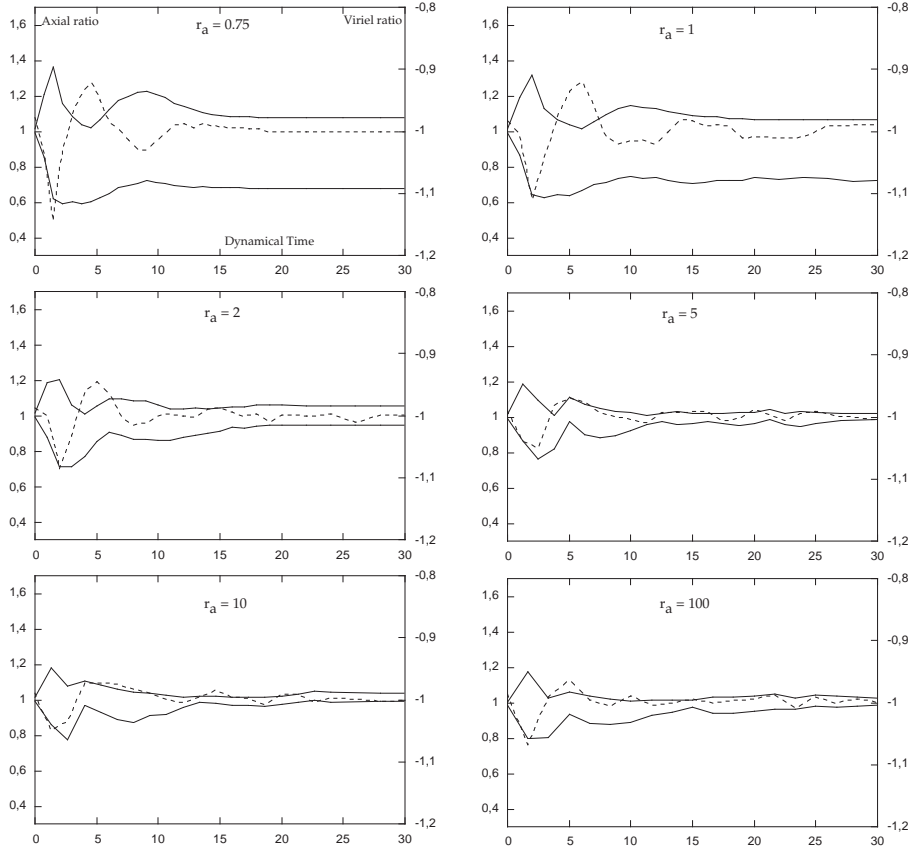

**Figure 6.** The axial (bold curves, left ordinates) and the virial (dashed curve, right ordinates) ratio vs dynamical time, for models with a polytropic index $n = 3.5$ and an anisotropy radius $r_a = 0.75$ (first diagram), $1, 2, 5, 10,$ and $100$.

### 4.2 Physical interpretation in terms of $\{g_1, E\}$ and $\{g_1, L^2\}$

The analytical developments presented in Section 2 concern the dynamics and the properties in the mean field approximation of the 6-variables DF $f$. In a N-body simulation we have access to the 6N-variables exact DF, $f^{(N)}(\mathbf{x}_1, \mathbf{v}_1, \cdots, \mathbf{x}_N, \mathbf{v}_N)$ which is a solution of the Liouville equation (for a detailed justification of the approximations required for re-constructing $f$ from $f^{(N)}$, see (Saslaw 1985))

$$\frac{\partial f^{(N)}}{\partial t} = \sum_{\alpha=1}^{N} \left( \nabla_x.(f^{(N)} \dot{\mathbf{x}}^\alpha) + \nabla_v.(f^{(N)} \dot{\mathbf{v}}^\alpha) \right) = 0. \tag{20}$$

Consequently, the Poisson brackets $\{g_1, E\}$ and $\{g_1, L^2\}$ appearing in the equation (14) defining the linear perturbation, are estimated in N-body simulations from statistical properties of the random variables $\epsilon_i$ and $\lambda_i$ defined for each particle $i$.

$$\epsilon_i \; : \; = \; \{g_1, E\}_{\mathbf{X}=\mathbf{X}_i}^{\mathbf{V}=\mathbf{V}_i} \; = \; (\nabla_x g_1.\mathbf{v})_i - (\nabla_v g_1.\nabla_x \psi)_i \; = \; \left(\frac{\partial E}{\partial \mathbf{v}}.d\mathbf{v}\right)_i + \left(\frac{\partial E}{\partial \mathbf{x}}.d\mathbf{x}\right)_i \; = \; dE_i\,, \tag{21}$$

and

$$\lambda_i \; : \; = \; \{g_1, L^2\}_{\mathbf{X}=\mathbf{X}_i}^{\mathbf{V}=\mathbf{V}_i} \; = \; \left(\nabla_x g_1.\nabla_v L^2\right)_i - \left(\nabla_v g_1.\nabla_x L^2\right)_i \; = \; \left(\frac{\partial L^2}{\partial \mathbf{v}}.d\mathbf{v}\right)_i + \left(\frac{\partial L^2}{\partial \mathbf{x}}.d\mathbf{x}\right)_i \; = \; dL_i^2\,. \tag{22}$$

$\epsilon_i$ and $\lambda_i$ are computed from positions and velocities of each particle $i$ at the initial time $t_0$ and at a time $t_1 = t_0 + \delta t$. In order to compare different models (defined by different $n$ and $r_a$) we tune $\delta t$ such that $\delta f / f = 0.01$. In all cases, we have $\delta t < T_d / 100$.

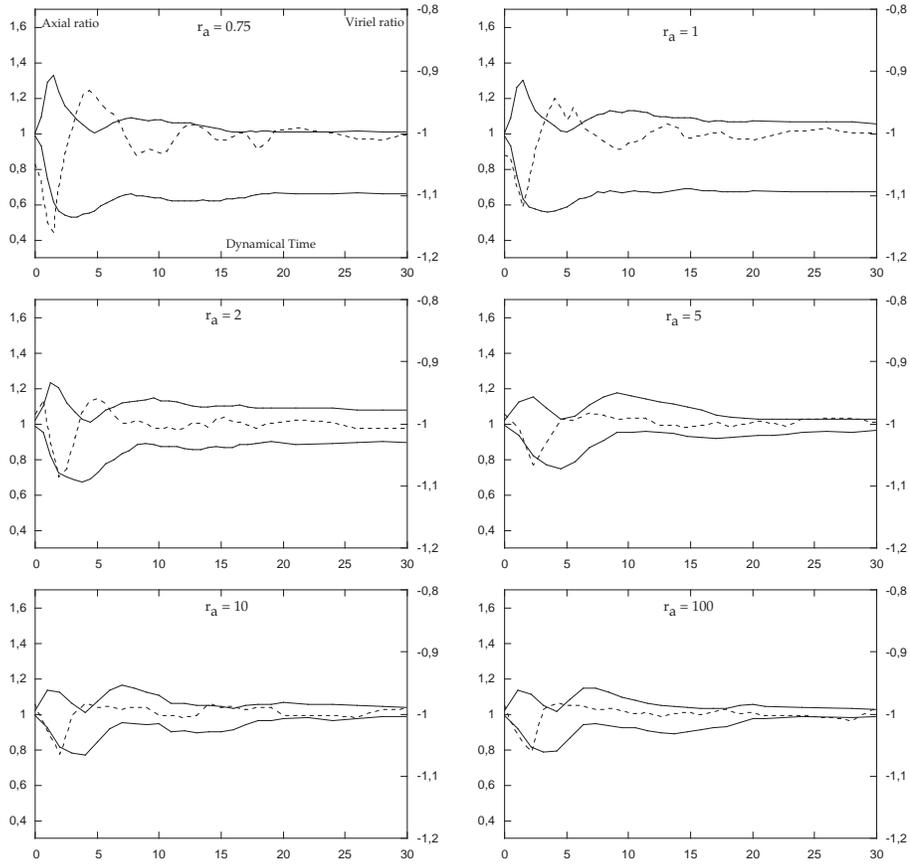

**Figure 7.** The axial (bold curves, left ordinates) and the virial (dashed curve, right ordinates) ratio vs dynamical time, for models with a polytropic index $n = 4$ and an anisotropy radius $r_a = 0.75$ (first diagram),$1, 2, 5, 10$, and $100$.

The variable $\epsilon_i$ is related to the perturbation in energy acting on the system (equation 21) and thus it is related to the stability of the system. In Fig. (9) , we plot for different models the fraction of $\epsilon_i$ in the numerical simulations at initial time with a negative value, i.e., the probability $P_{nb}(\epsilon)$ for $\epsilon$ to be negative.

The variable $\lambda_i$ is related to the anisotropic component of the linear variation of the DF of the system (equation 22). We have shown (Section 2, (Perez 1995), Paper I) that the systems have particular stability properties against all preserving perturbations, i.e., the perturbations generated by a $g_1$ such that $\{g_1, L^2\} = 0$. But except the purely isotropic stable case, there exist no physical systems which are submitted only to preserving perturbations. In the opposite case, a fully anisotropic system with all components evolving only on radial orbits, is unstable. Consequently, in order to characterize the preserving nature of the perturbations acting on the system, we suggest to use the distribution of $\lambda_i$ around their vanishing mean value. In order to discriminate between the preserving and non-preserving nature of perturbations we must take into account not only how weakly the $\lambda_i$ are scattered around their mean, but also how highly they are peaked at their mean. We thus compute the statistical Pearson index of the random variable $\lambda$ (hereafter $P_\lambda$). This parameter is both a shape and a scattering parameter. A flattened distribution (platikurtic) of a random variable is characterized by a small or negative Pearson index. A distribution which is both concentrated around its mean value and which is highly peaked (leptokurtic) has a large Pearson index (Calot 1973). However, as we are interested only in the distribution of $\lambda_i$ around their mean value and, in order to eliminate some aberrant contributions, we define a truncated variable

$$\overline{\lambda_i} = \begin{cases} \lambda_i & \text{when} \quad |\lambda_i| \leq 3\sigma_\lambda \\ \underline{\lambda} & \text{when} \quad |\lambda_i| > 3\sigma_\lambda \end{cases}$$

where $\sigma_\lambda$ is the standard deviation of the random variable $\lambda_i$ and $\underline{\lambda}$ is the mean of $\overline{\lambda_i}$. We compute the statistical Pearson index of the variable $\overline{\lambda_i}$

$$P_{\overline{\lambda}} = \frac{\sum_{i=1}^N \left(\overline{\lambda_i} - \underline{\lambda}\right)^4}{\left(\sum_{i=1}^N \left(\overline{\lambda_i} - \underline{\lambda}\right)^2\right)^2} - 3, \tag{23}$$

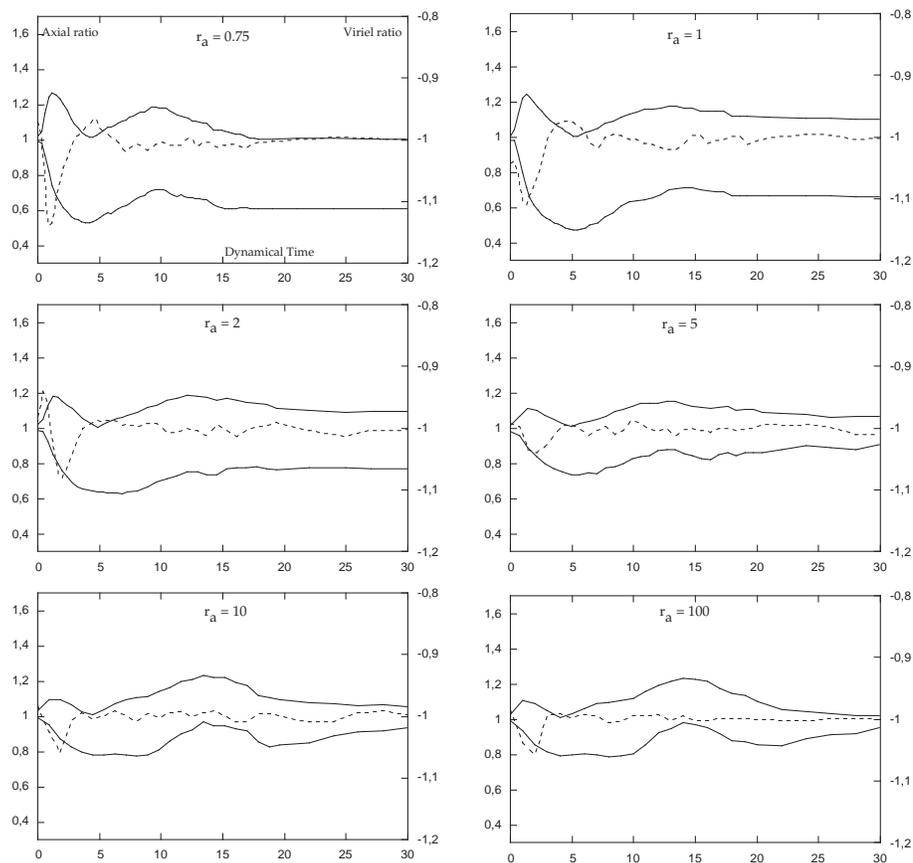

**Figure 8.** The axial (bold curves,left ordinates) and the virial (dashed curve, right ordinates) ratio vs dynamical time, for models with a polytropic index $n = 4.5$ and an anisotropy radius $r_a = 0.75$ (first diagram), $1, 2, 5, 10$, and $100$.

In Fig. (9), we plot $P_{\frac{1}{\lambda}}$ for different models. The error bars both for $P_{\frac{1}{\lambda}}$ and $P_{nb}(\epsilon)$ are calculated from 6 numerical simulations of a given model. The bars represent $\pm \sigma$ variations (i.e., 95% of the values for this non-gaussian variable).

For any polytropic index $n$, the respective behaviors of $P_{nb}(\epsilon)$ and $P_{\frac{1}{\lambda}}$ are similar. As a matter of fact, the strongly anisotropic initial systems with $r_a \lesssim 2$ which are morphologically unstable (Section 4.1, Fig. 5-6-7), are characterized by a large $P_{nb}(\epsilon)$,

$$P_{nb}(\epsilon) \gtrsim 20\% \,, \tag{24}$$

and by a small $P_{\frac{1}{\lambda}}$

$$P_{\frac{1}{\lambda}} \lesssim 2.5 \,. \tag{25}$$

On the contrary, the median and weakly anisotropic initial systems ($r_a \gtrsim 2$), which are morphologically stable (Section 4.1, Fig. 5-6-7), are characterized by a small $P_{nb}(\epsilon)$,

$$P_{nb}(\epsilon) \lesssim 20\% \,, \tag{26}$$

and a large $P_{\frac{1}{\lambda}}$,

$$P_{\frac{1}{\lambda}} \gtrsim 2.5 \,. \tag{27}$$

Moreover, these two classes of systems are clearly separated with respect to the logarithmic scale used for $r_a$.

## 5 UNIVERSALITY OF OUR PREDICTIONS AND CONCLUSIONS

In order to generalize to all collisionless self-gravitating spherical systems the stability criterion – defined in terms of the parameters $P_{nb}(\epsilon)$ and $P_{\frac{1}{\lambda}}$ – which has been obtained for the Ossipkov-Merritt models (equation (24)-(27)), we have generated four independant spherical models. These models do not follow an Ossipkov-Merritt distribution (Section 3.1). They are virialized and present an anisotropy in the velocity space. We show in Table 1 the physical characteristics (size, dynamical time) of these different configurations. Since all these models are spherical and anisotropic, their DF depends both on $E$ and $L^2$. We plot the iso-contour of this function in the $E - L^2$ plane, thus forming the so-called Lindblad diagram of the system (see (Merritt 1985a), (Lindblad 1933)). For comparison, we have plotted in Fig. 10 a set of such diagrams for the Ossipkov-Merritt models with $n = 4.5$, $r_a = 1, 2, 10$ and 100. We show in Fig. 11 the Lindblad diagram of the systems described in Table 1.

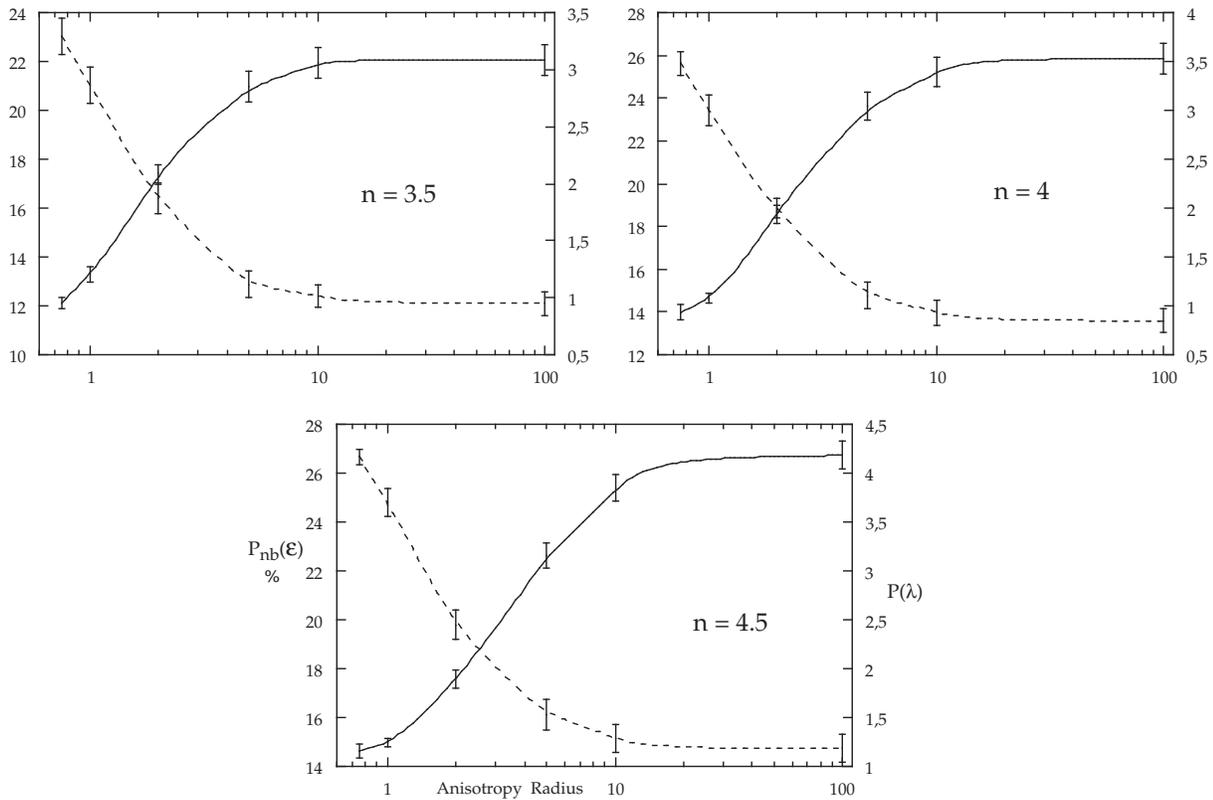

**Figure 9.** Probability $P_{nb}(\epsilon)$ for $\epsilon$ to be negative (dashed curves, left ordinates) and Pearson index of the random variable $\lambda$ (plain curves, right ordinates) vs anisotropy radius, for models with a polytropic index $n = 3.5$, 4, and 4.5, respectively.

We have considered four models in order to illustrate the three classes of foreseable spherical systems. : quasi-isotropic models (M1 and M3), circular anisotropic models (M4) and a non Ossipkov-Merritt radial anisotropic system (M2).

Models 1 and 3 are weakly anisotropic. They are similar to the Ossipkov-Merrit systems (respectively with $n = 4.5$, $r_a = 10$ and $n = 4.5$, $r_a = 100$). Model 2 represents a system with particles mainly on radial orbits. Its DF is highly peaked around $L^2 = 0$. An analysis in velocity space of model 4 shows that this model presents a strong circular orbit anisotropy.

In Fig. (12), we present the evolution of the axial and virial ratios for models 1 to 4. Only the model 2 is morphologically unstable, contrary to model 1 which is very stable. We have computed for these models the parameters $P_{nb}(\epsilon)$ and $P_{\overline{\lambda}}$. The stability criteria (equations (24)-(27)) are confirmed. For the unstable model 2, parameter $P_{nb}(\epsilon)$ is larger than 20% (Table 1) and parameter $P_{\overline{\lambda}}$ is smaller than 2.5 (Table 1). The very stable model 1 is characterized by a low $P_{nb}(\epsilon)$ ($= 18.75\% < 20\%$) and by a very large $P_{\overline{\lambda}}$ ($= 8.26 > 2.5$) (Table 1). We are thus able to generalize our stability criteria from the previous section to all collisionless self-gravitating spherical systems:

• All anisotropic collisionless self-gravitating spherical systems with parameter $P_{nb}(\epsilon)$ smaller than 20% are stable. As we do not have analytical suggestions for the opposite case, we do not propose any conclusions on the instability of such systems from parameter $P_{nb}(\epsilon)$.

• The dynamical evolution of all anisotropic collisionless self-gravitating spherical systems with $P_{\overline{\lambda}} < 2.5$ is dominated by non-preserving perturbations (Section 2). Such systems are unstable.

**Table 1.** Physical characteristic of used general anisotropic spherical systems

|         | $R_{\frac{1}{2}}$ | $T_d$ | $P_{nb}(\epsilon)$ | $P_{\frac{}{\lambda}}$ |
|---------|------|-------|--------|-------|
| Model 1 | 2.69 | 4,42  | 18,75  | 8,26  |
| Model 2 | 9.94 | 31,35 | 22,82  | 2,12  |
| Model 3 | 3.27 | 5,94  | 17,46  | 2,93  |
| Model 4 | 3.18 | 5,68  | 18,04  | 3,42  |

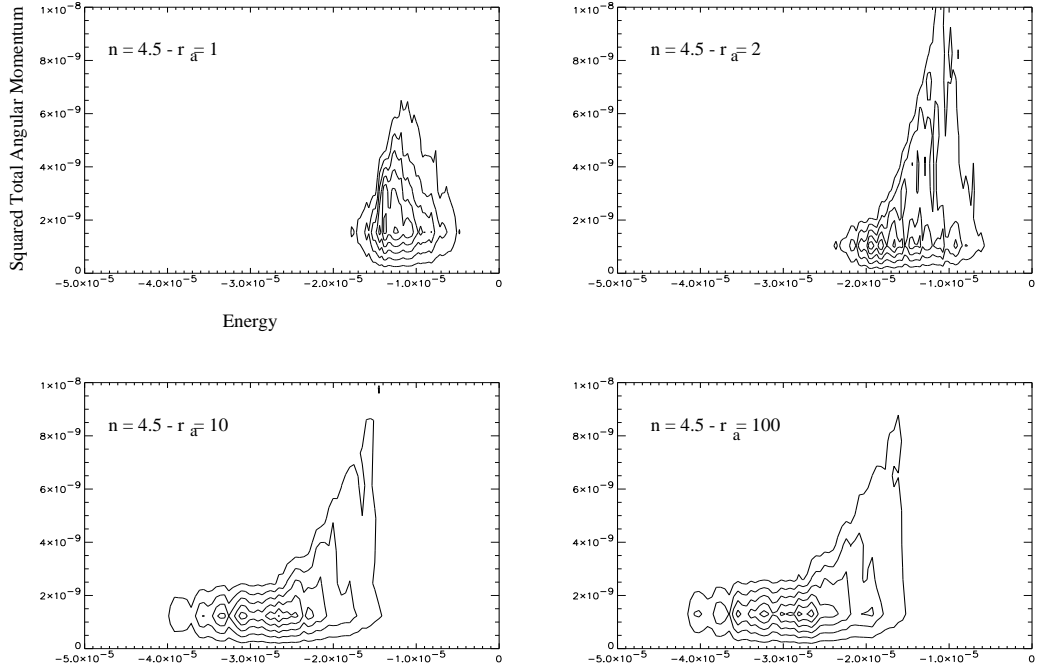

**Figure 10.** Lindblad diagram for polytropic Ossipkov-Merritt models with $n = 4.5$ $r_a = 1, 2, 10$, and $100$.

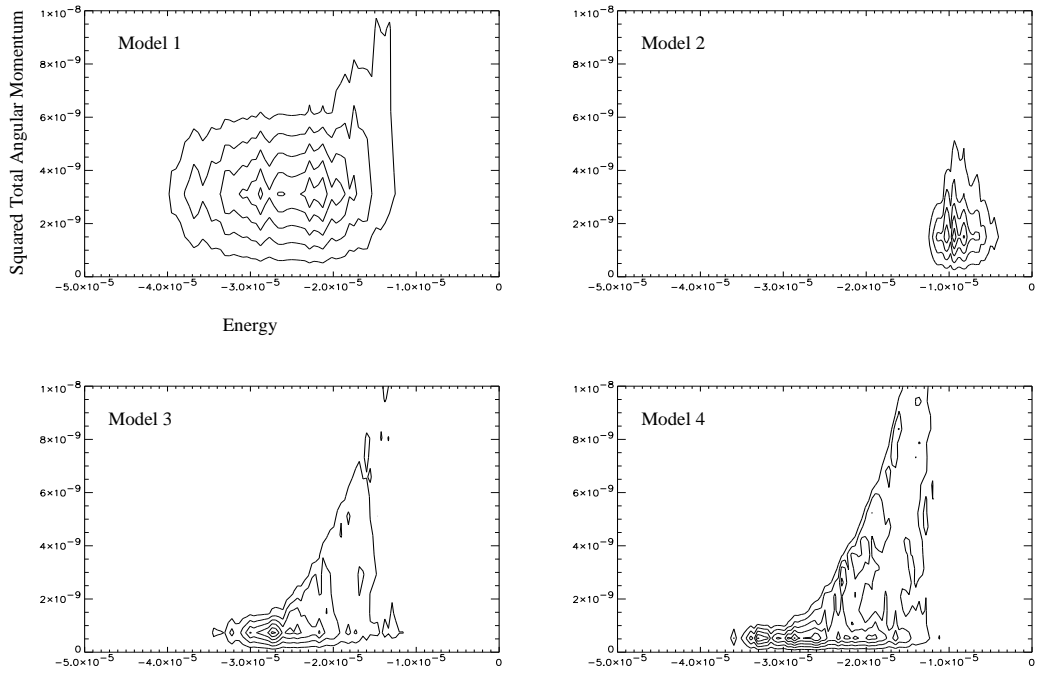

**Figure 11.** Lindblad diagram for models 1,2,3 and 4.

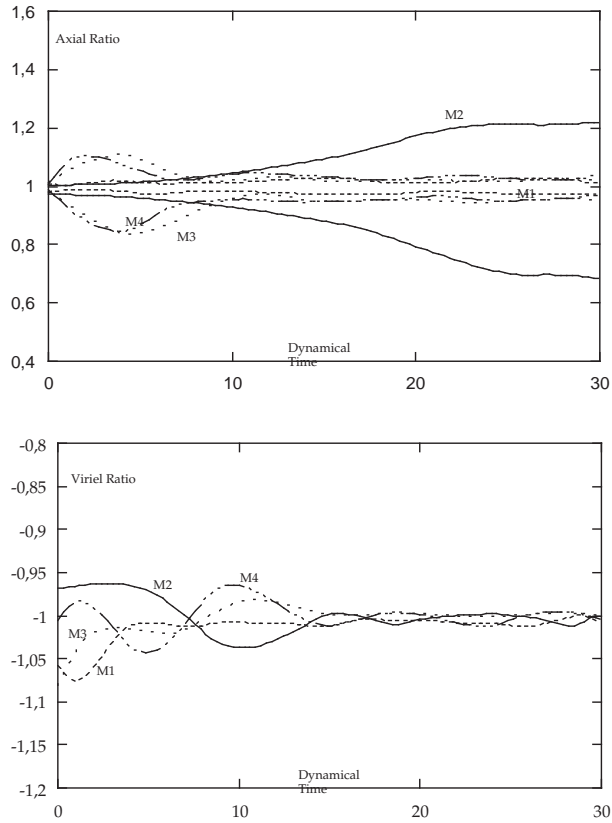

**Figure 12.** Axial ratio (top diagram) and virial ratio (bottom diagram) vs dynamical time for models 1,2,3 and 4.

**Table A1.** Physical characteristics for models with $n = 3.5,, 4, 4.5$ and $r_a = 0.75, 1, 2, 5, 10$ and $100$

| $n = 3.5$ | $T_d$ | $R_{\frac{1}{2}}$ | $P_{nb}(\epsilon)$ | $P(\lambda)$ |
|---|---|---|---|---|
| $r_a = 0.75$ | 5.455 ±0.062 | 3.099 ±0.023 | 23.04 ±0.72 | 0.953 ±0.053 |
| $r_a = 1$ | 5.127 ±0.057 | 2.973 ±0.022 | 21.03 ±0.73 | 1.205 ±0.064 |
| $r_a = 2$ | 4.157 ±0.039 | 2.585 ±0.016 | 16.38 ±0.61 | 2.088 ±0.076 |
| $r_a = 5$ | 3.238 ±0.022 | 2.189 ±0.010 | 12.87 ±0.55 | 2.85 ±0.13 |
| $r_a = 10$ | 3.043 ±0.020 | 2.0997 ±0.0095 | 12.38 ±0.46 | 3.06 ±0.13 |
| $r_a = 100$ | 2.973 ±0.020 | 2.0676 ±0.0097 | 12.10 ±0.48 | 3.082 ±0.13 |
| $n = 4$ | $T_d$ | $R_{\frac{1}{2}}$ | $P_{nb}(\epsilon)$ | $P(\lambda)$ |
| $r_a = 0.75$ | 8.720 ±0.096 | 4.237 ±0.031 | 25.63 ±0.60 | 0.935 ±0.082 |
| $r_a = 1$ | 8.257 ±0.087 | 4.085 ±0.029 | 23.45 ±0.72 | 1.079 ±0.044 |
| $r_a = 2$ | 6.473 ±0.037 | 3.472 ±0.013 | 18.74 ±0.57 | 1.962 ±0.065 |
| $r_a = 5$ | 4.503 ±0.037 | 2.727 ±0.015 | 14.77 ±0.64 | 3.04 ±0.15 |
| $r_a = 10$ | 4.015 ±0.033 | 2.526 ±0.014 | 13.98 ±0.60 | 3.40 ±0.12 |
| $r_a = 100$ | 3.835 ±0.031 | 2.450 ±0.013 | 13.59 ±0.55 | 3.53 ±0.16 |
| $n = 4.5$ | $T_d$ | $R_{\frac{1}{2}}$ | $P_{nb}(\epsilon)$ | $P(\lambda)$ |
| $r_a = 0.75$ | 17.09 ±0.21 | 6.634 ±0.053 | 26.67 ±0.38 | 1.161 ±0.070 |
| $r_a = 1$ | 16.26 ±0.19 | 6.418 ±0.050 | 24.79 ±0.56 | 1.241 ±0.042 |
| $r_a = 2$ | 12.56 ±0.12 | 5.403 ±0.035 | 19.78 ±0.60 | 1.898 ±0.094 |
| $r_a = 5$ | 7.206 ±0.081 | 3.730 ±0.028 | 16.12 ±0.61 | 3.16 ±0.13 |
| $r_a = 10$ | 5.655 ±0.062 | 3.174 ±0.023 | 15.14 ±0.57 | 3.85 ±0.13 |
| $r_a = 100$ | 5.064 ±0.051 | 2.949 ±0.020 | 14.73 ±0.57 | 4.18 ±0.14 |

## APPENDIX A: STATISTICS OF SIMULATIONS

We have randomly choosen the position and the velocity for 8192 particles from the DF. Six independent simulations have been carried out for each value of $n$ and $r_a$. We present in the following tables the physical properties and the stability parameters calculated for a large class of generalized polytropic Ossipkov-Merritt models. The dynamical time is the ratio between the typical size of the system and the modulus of median velocity of a particle. $R_{\frac{1}{2}}$ represents the radius of the sphere containing half of the system total mass. $P_{nb}(\epsilon)$ represents the probability for the $\{g_1, E\}$ calculated for each particle to be negative (expressed in percent). Finally, $P(\lambda)$ represents the Pearson index of the distribution of $\{g_1, L^2\}$. The number indicates the mean value of the quantity over the 6 simulations, and the error represents an interval of two standard deviations of the quantity. Units are such that the gravitationnal constant, the total mass of the system and the initial value of the relative gravitationnal potential are unity.